\documentclass{article}

\usepackage{amsmath,amsthm,amssymb}

\title{%
  Mechanistic Behavior of
  Single-Pass Instruction 
  Sequences\footnotetext{This work has been carried out within and funded by the NWO project "Thread Algebra for Strategic Interleaving".}}
\author{%
  Jan A.\ Bergstra \and
  Mark B.\ van der Zwaag\\[2ex]
  {\footnotesize Section Software Engineering,
  Informatics Institute, University of Amsterdam}
  }
\date{}%\today, Version 5}

\theoremstyle{definition}
\newtheorem{example}{Example}
\newtheorem{fact}{Fact}
\newtheorem{definition}{Definition}

\newcommand{\delay}{\ensuremath{\sigma}}
\newcommand{\funabs}{\ensuremath{\mathit{fa}}}
\newcommand{\mleq}{\ensuremath{\sqsubseteq_\delay}}
\newcommand{\mlt}{\ensuremath{\sqsubset_\delay}}
\newcommand{\BTA}{\ensuremath{\textrm{BTA}}} 
\newcommand{\BTAs}{\ensuremath{\textrm{BTA}_{\delay}}} 
\newcommand{\di}{\ensuremath{\mathsf{D}}}
\newcommand{\st}{\ensuremath{\mathsf{S}}}
\newcommand{\tr}{\ensuremath{{\mathtt{true}}}}
\newcommand{\fa}{\ensuremath{{\mathtt{false}}}}
\newcommand{\pcc}[3]{\ensuremath{#2 \unlhd #1 \unrhd #3}}
\newcommand{\Act}{\ensuremath{\mathit{A}}}
\newcommand{\BI}{\ensuremath{\mathcal{A}}}
\newcommand{\extr}[1]{\ensuremath{|#1|}}
\newcommand{\mextr}[1]{\ensuremath{|#1{|}^{\delay}}}

\newcommand{\Tin}{\ensuremath{\mathalpha{!}}} % Termination
\newcommand{\Pt}[1]{\ensuremath{{+}#1}} % Positive test
\newcommand{\Nt}[1]{\ensuremath{{-}#1}} % Negative test
\newcommand{\Fj}[1]{\ensuremath{{\#}#1}} % Forward jump
\newcommand{\Rep}[1]{\ensuremath{{(}#1{)^{\omega}}}} % Repeat
\begin{document}
  \maketitle

\begin{abstract}
Earlier work on program and thread algebra
detailed the functional, observable behavior
of programs under execution.
In this article we add the modeling of
unobservable, \emph{mechanistic}
processing, in particular processing due to
jump instructions.
We model mechanistic processing preceding
some further behavior as a \emph{delay} of that behavior;
we borrow a unary delay operator
from discrete time process algebra.
We define a mechanistic improvement ordering
on threads and observe that some threads
do not have an optimal implementation.
\end{abstract}

\section{Introduction}
We recall from~\cite{BL02} the notion
of an instruction sequence and its (functional) thread extraction.
Processing cost constitutes a 
non-functional aspect of instruction sequences.
Below we will define a version of thread extraction that takes
the cost of jump instructions into account.
The simplest intuition for this cost is processing time,
under the assumption that the instruction is used as a
machine code which is not further compiled before processing on a
suitable machine.
The presence of a jump induces a delay which is independent of
the size of the jump.
We call the thread extracted from an instruction sequence while
taking cost of jumps into account its mechanistic behavior.
The mechanistic behavior reflects some non-functional properties
that arise from the execution mechanism. 
Given the definition of a mechanistic behavior we can
define when an implementation of a thread improves
another implementation and when an implementation is
(locally) optimal or globally optimal.
After clarifying the definitions with a number of examples
it is shown that some implementable threads have
no optimal implementations: each implementation can be improved.

Mechanistic behavior is an essential ingredient for a theory of
instruction sequences.
Indeed compilation and code generation can often be viewed as
steps transforming an instruction sequence into
a functionally equivalent one with
an improved mechanistic behavior. 
This paper provides an approach to the quantification of
non-functional aspects of instruction sequences
from first principles which might eventually
enable a useful analysis of compilation methods as well
as a principled investigation and explanation of
what constitutes the best possible design of
a set of instructions for writing machine programs.

Following~\cite{BL02} we use single pass instruction sequences
for two reasons: 
it gives rise to a convenient algebra of instruction sequences
and it allows to simplify the definition of thread extraction
to a bare minimum.
We refer to~\cite{PZ06} for further introductory
information on instruction sequences and threads.

\section{Instruction Sequences}
Program algebra (PGA, for ProGram Algebra, see~\cite{BL02})
provides a framework for 
the understanding of imperative sequential programming.
Starting point is the perception of a program
as an expression of an 
\emph{instruction sequence} --- a
possibly infinite sequence
\[ u_1;u_2;u_3;\ldots
\]
of \emph{primitive} instructions $u_i$.

Given a set \BI\ of \emph{basic} instructions,
the primitive instructions of PGA are the following:

\begin{description}

\item[{Basic void instruction}.]
  All elements of \BI, written, typically, as 
  $a,b,\ldots$,
  can be used as \emph{basic void instructions}. 
  These are regarded as indivisible units and
  execute in finite time.
  The associated behavior may modify a state.

\item[{Termination instruction}.]
  The termination instruction \Tin\
  yields successful termination of the execution.
  It does not modify a
  state, and it does not return a boolean value.

\item[{Basic test instruction}.]
  A basic instruction $a\in\BI$
  is viewed as a request to the
  environment, and it is assumed that upon its execution
  a boolean value (\tr\ or \fa)
  is returned that may be used for subsequent program control.
  For each element $a$ of \BI\ there is
  a \emph{positive test} instruction \Pt{a} and 
  a \emph{negative test} instruction \Nt{a}.
  When a positive test is
  executed, the state is affected according to $a$,
  and in case \tr\ is returned, the remaining sequence of
  actions is performed. 
  If there are no remaining instructions, inaction occurs.
  In the case that \fa\ is returned,
  the next instruction is skipped and execution
  proceeds with the instruction following the skipped one.
  If no such instruction exists, inaction occurs.
  Execution of a negative test is the same, except that 
  the roles of \tr\ and \fa\ are interchanged.
  
\item[{Forward jump instruction}.]
  For any natural number $k$, the instruction \Fj{k} 
  denotes a jump of length $k$ and $k$ is called the counter 
  of this instruction.
  If $k=0$, this jump is to the instruction itself and
  inaction occurs (one can say that \Fj{0} 
  \emph{defines} divergence,
  which is a particular form of inaction).
  If $k=1$, the instruction skips itself, and execution
  proceeds with the subsequent instruction if available,
  otherwise inaction occurs.
  If $k>1$, the instruction \Fj{k} skips itself and
  the subsequent $k-1$ instructions.
  If there are not that many instructions left in the 
  remaining part of the program, inaction occurs.
\end{description}

In PGA, a \emph{program}
is an expression (in a programming language)
that represents an instruction sequence.
In~\cite{BL02}, a hierarchy of programming
languages is built, with languages
containing constructs of increasing complexity
such as labels and goto's,
conditionals and while-loops, etc.
Programs in these languages 
can always be projected to an expression at a basic 
level where it can be mapped to
an instruction sequence directly. 
More specifically, at this basic level
PGA allows to build programs from the
primitive instructions listed above by means of two composition 
operators.

First, we have
instruction sequence concatenation, written $X;Y$
for instruction sequences $X$ and $Y$.
Concatenation is supposed to 
be associative, so that its parentheses are usually omitted.

The second PGA operator is repetition, written $X^{\omega}$,
representing the infinite concatenation
$X;X;X;\ldots$.
Repetition unfolds in the following 
way: $X^{\omega}= X;X^{\omega}$,
and
if $X$ is an infinite instruction sequence already,
we will use $X^{\omega} = X$.

Below we will restrict attention to instruction sequences
that can be written in PGA notation.
This means that instruction
sequences will be either finite or eventually periodic.

\section{Functional and Mechanistic Behaviors}
The \emph{execution} of an instruction sequence
is single-pass: the instructions are visited in order and 
are dropped after having been executed.
Execution of a basic instruction is interpreted as
a request to the execution environment: the
environment processes the request and replies with
a Boolean value.
This has led to the modeling of the functional behavior 
of instruction sequences as threads, that is,
as elements of Basic Thread Algebra (BTA).
An interpretation mapping $\extr{\_}$
from instruction sequences to threads is
given in Section~\ref{sec:thrextr}.
This interpretation is
called \emph{thread extraction}.

Based on a set $\Act$ of \emph{actions}, which
will be used to interpret basic instructions,
BTA has the following constants and operators:
\begin{itemize}
\item the \emph{termination} constant $\st$,
\item the \emph{deadlock} or \emph{inaction} constant $\di$,
\item for each $a\in A$, a binary 
  \emph{postconditional composition}
  operator $\_  \unlhd a \unrhd \_$.
\end{itemize}
We use \emph{action prefixing} $a \circ P$ as an abbreviation for
$P \unlhd a \unrhd P$ and take $\circ$ to bind strongest.
Furthermore, for $n\geq 1$ 
we define $a^n\circ P$ by $a^1\circ P=a\circ P$ and 
$a^{n+1}\circ P=a\circ (a^n\circ P)$.

The operational intuition
is that each action represents a command which is to be
processed by the execution environment of the thread.
The processing of a command may involve a change of
state of this environment. 
At completion of the processing of the command,
the environment
produces a reply value \tr\ or \fa.
The thread $P \unlhd a \unrhd Q$
proceeds as $P$ if the processing of $a$ yields \tr,
and it proceeds as $Q$ if the processing of
$a$ yields \fa.

Every thread in BTA is finite in the sense that there
is a finite upper bound to the number of consecutive 
actions it can perform.
BTA has a completion which comprises also the infinite threads.
We interpret instruction sequences, which may be infinite,
in this completion.\footnote{
	We omit the mathematical
	details of this construction because these are not essential for
	an understanding of the paper.
	In~\cite{BL02} the completion has been
	worked out in terms of projective limits,
	but other constructions are possible as well.
	The formalization of infinite objects on the basis of 
	finite ones can be done in different ways and the development
	here is not specific for a particular choice
	of that formalization.
}

\paragraph{Mechanistic Threads.}
To model non-functional aspects of
the behavior of instruction sequences
(in particular the processing of jumps),
we extend BTA with a unary \emph{delay operator} $\delay$
taken from relative discrete time process
algebra~\cite{BB96,BM98}.
Write \BTAs\ for this extension.
Notation:
define $\delay^0(P)=P$ and
$\delay^{n+1}(P)=\delay(\delay^n(P))$.\footnote{
	In relative discrete time process algebra,
	the delay operator defers the contained
	behavior to the next time slice.
	It is assumed that time progresses in slices
	of equal length, and that the execution of
	actions does not take time: actions are 
	executed within a time slice.
	In our sequential setting case  such assumptions
	are not needed.
	In fact an `effort' or `cost' interpretation
	is just as valid as a `time' interpretation.
	But in any interpretation we do assume
	that the size of one delay is fixed, and that
	delays can be added: $\delay^{n+1}(P)$ always
	puts a strictly larger delay on $P$ than $\delay^n(P)$
	unless $P=\di$.
}

We call the elements of \BTA\ \emph{functional} threads,
and the elements of \BTAs\ \emph{mechanistic} threads.
For example,
\[ \delay(a\circ\st)
\]
is a mechanistic thread defining the
functional behavior $a\circ\st$ preceded
by one delay.

Observe that $\BTA\subset\BTAs$, so
any functional thread is also a mechanistic thread.
To take this further, 
we define the functional behavior
of a mechanistic thread
as the thread that is obtained
if we remove all delays.
The \emph{functional abstraction operator} 
$\funabs(\_)$
does just this:
\begin{align*}
\funabs(\st) &= \st\\
\funabs(\di) &= \di\\
\funabs(\pcc aPQ) &= \pcc a{\funabs (P)}{\funabs(Q)}\\
\funabs(\delay(P)) &= \funabs(P)
\end{align*}
So, for any mechanistic thread $P\in\BTAs$,
the functional thread $\funabs(P)\in\BTA$
stands for the functional behavior of $P$.
Two mechanistic threads $P,Q\in\BTAs$
are \emph{functionally equivalent},
notation $P\sim_f Q$,
if they have the same functional behavior:
we define
\[ P\sim_f Q\quad\text{iff}\quad \funabs(P) = \funabs(Q).
\]

\paragraph{Mechanistic Improvement.}
The obvious question is now
how to compare distinct mechanistic threads
that are functionally equivalent.

For example,
we find that
\begin{align*}
\delay^{452}(P)&\sim_f \delay^{3}(P),\\
\pcc a {\delay(P)}{Q} & \sim_f \pcc a {P}{\delay(Q)}.
\end{align*}
In the first case, 
the right-hand side $\delay^3(P)$
yields the same functional behavior
as the left-hand side using far fewer delays
preceding the execution of $P$.
We call $\delay^3(P)$ a 
\emph{mechanistic improvement}
of $\delay^{452}(P)$ (an ordering is defined formally below).
A mechanistic improvement is
viewed as a more efficient way 
to obtain a (desired) functional behavior.
The threads in the second example above
cannot be compared in this way:
the delays that are visible here in their
respective true and false branches occur
under different circumstances (execution histories).

The mechanistic improvement ordering
$\mleq$ is defined as follows.
\begin{align*}
P &\mleq P,\\
P &\mleq \delay(P),\\
\delay(\di) &\mleq \di,
\end{align*}
and
\[ P\mleq P',\ Q\mleq Q'\quad\text{imply}\quad
   \pcc a{P}{Q} \mleq \pcc a{P'}{Q'}.
\]
If $P\mleq Q$, we say that $P$ is a mechanistic 
improvement of $Q$.
Further write $P\mlt Q$ if
$P$ is a strict mechanistic improvement of $Q$
(so $P\mleq Q$ and $P\neq Q$).

An obvious observation is that
mechanistic improvements yield the same
functional behavior:
\[ P\mleq Q\quad\text{implies}\quad
   P\sim_f Q.
\]
As seen in the example above,
functionally equivalent mechanistic threads
need not be comparable by the
mechanistic improvement ordering:
\[ P\sim_f  Q\quad\text{does not imply}\quad
   (P\mleq Q~\text{or}~Q\mleq P).
\]

\section{Thread Extraction}
\label{sec:thrextr}
The
\emph{thread extraction operator} \extr{\_} assigns a
possibly infinite 
BTA thread to a PGA instruction sequence.
The resulting thread models the functional behavior
of the sequence:
basic instructions are interpreted
as (observable) actions, while
the interpretation of jump instructions
is made part of the extraction.

Thread extraction is defined by the following equations 
(where $a$ ranges over basic instructions,
$u$ over primitive instructions,
and $X$ over non-empty sequences):
\begin{align*}
\extr{X} &= \extr{X;\Fj 0} \qquad\text{if }X\text{ is finite}\\
\extr{\Tin; X} &= \st \\
\extr{a;X} &= a\circ \extr{X}\\
\extr{\Pt a; u; X} &= \pcc{a}{\extr{u;X}}{\extr{X}}\\
\extr{\Nt a; u; X} &= \pcc{a}{\extr{X}}{\extr{u;X}}\\
\extr{\Fj 0; X} &= \di\\
\extr{\Fj 1; X} &= \extr{X}\\
\extr{\Fj{k+2}; u; X} &= \extr{\Fj{k+1}; X}
\end{align*}
Observe that we interpret basic instructions as actions;
we use the same symbol to denote both the instruction
and its interpretation.

The functional interpretation of instruction sequences
defined above abstracts from the (mechanistic) 
processing of jump instructions.
For example, the sequences
\[ \Fj 1;\Fj 1; a;\Tin
\]
and
\[ a;\Tin
\]
yield the same functional behavior, namely,
the thread $a\circ\st$.
Still, execution of the first sequence may require more
time and effort because
of the processing of the jump instructions.
We define an alternative thread extraction
operator that takes the mechanistic aspect of
behavior into account.
We assume that the processing of
a jump instruction $\Fj k$, irrespective of the value
of $k$, results in one \emph{delay}
of the subsequent behavior.

The mechanistic thread extraction \mextr{\_}
is defined by the following equations:
\begin{align*}
\mextr{X} &= \mextr{X;\Fj 0} \qquad\text{if }X\text{ is finite}\\
\mextr{\Tin; X} &= \st \\
\mextr{a;X} &= a\circ \mextr{X}\\
\mextr{\Pt a; u; X} &= \pcc{a}{\mextr{u;X}}{\mextr{X}}\\
\mextr{\Nt a; u; X} &= \pcc{a}{\mextr{X}}{\mextr{u;X}}\\
\mextr{\Fj 0; X} &= \di\\
\mextr{\Fj 1; X} &= \delay(\mextr{X})\\
\mextr{\Fj{k+2}; u; X} &= \mextr{\Fj{k+1}; X}
\end{align*}

\begin{fact}
The functional thread extraction
\extr{X} of sequence $X$ equals
the functional abstraction of the 
mechanistic interpretation of $X$:
\[ \extr X = \funabs(\mextr X)
\]
for all sequences $X$.
\end{fact}

\begin{example}

For both $(\Fj 1;a)^\omega$ and
$(\Fj 2;\Fj 1; a)^\omega$,
mechanistic thread extraction yields the thread
$P$ defined recursively by $P=\delay(a\circ P)$.

The mechanistic interpretation 
of $(\Fj 1;\Fj 1; a)^\omega$ yields
$P=\delay^2(a\circ P)$, and for
$(\Fj 2; a)^\omega$ it yields $P=\delay(P)$.
\end{example}

\section{Implementations}

\begin{definition}[Mechanistic pre-extraction]
Instruction sequence $X$ is a
\emph{mechanistic pre-extraction}
of thread $P$, if
\[ P = \mextr X.
\]
\end{definition}

A mechanistic pre-extraction of a thread $P$
is a particular implementation of the behavior $P$
(in fact, it is a particularly efficient implementation,
see Fact~\ref{fact:x} below),
where an implementation is defined as follows.

\begin{definition}[Implementation]
Instruction sequence $X$ is an
\emph{implementation}
of thread $P$, if
\[ P\mleq\mextr X.
\]
\end{definition}

\begin{fact}\label{fact:xyz}
Not all (implementable)
threads have a mechanistic pre-extraction.
For example, consider
\[ P = \pcc a{b\circ\st}{c\circ\st}.
\]
It is not difficult to see that $P$ does
not have a mechanistic pre-extraction:
any implementation will
contain at least one jump instruction
leading to extractions containing delays
not present in $P$.
The sequence
\[
   X = \Pt{a};\Fj 3; c;\Tin;b;\Tin
\]
with
\[
  \mextr X = \pcc{a}{\delay(b\circ\st)}{c\circ\st}
\]
is an implementation of $P$.
\end{fact}

We compare implementations of a thread
by their respective mechanistic extractions.
That is, if $X$ and $Y$ are implementations
of $P$, then we say that $X$ is a
mechanistic improvement of $Y$
if \mextr{X} is a
mechanistic improvement of \mextr{Y},
that is, if $\mextr X\mleq\mextr Y$. 

\begin{definition}[Optimal implementation]
Instruction sequence $X$ is an
\emph{optimal implementation}
of thread $P$, if
\[ P\mleq\mextr X,
\] 
and for no other instruction sequence $Y$ 
that implements $P$ we have
$\mextr{Y} \mlt\mextr{X}$.
\end{definition}

\begin{definition}[Globally optimal implementation]
Instruction sequence $X$ is a
\emph{globally optimal implementation}
of thread $P$, if
\[ P\mleq\mextr X,
\] 
and for each other instruction sequence $Y$ 
that implements $P$ we have
$\mextr{X} \mleq\mextr{Y}$.
\end{definition}

\begin{fact}
\label{fact:x}
If a thread $P$ has a mechanistic pre-extraction $X$,
that is, $\mextr X = P$,
then this $X$ is a globally optimal implementation
of $P$.
\end{fact}

\begin{example}
We find that the sequences
\[ X= \Fj 1;\Fj 1; a;\Tin
   \quad\text{and}\quad
   Y = \Fj 1; a;\Tin
\]
yield the respective mechanistic
extractions $\delay^2(a\circ\st)$ and $\delay(a\circ\st)$.
Both $X$ and $Y$ are implementations of
$P=a\circ\st$,
and $Y$ is a mechanistic improvement of $X$.
Observe that
the mechanistic pre-extraction $Z= a;\Tin$ of $P$
further improves $Y$,
and that $Z$ is a globally optimal implementation of $P$.
\end{example}

\begin{example}
Consider thread $P$ defined by $P = \pcc a{P}{b\circ \st}$.
Then $X = (\Pt a; \Fj 3;b ; \Tin)^{\omega}$ is an optimal 
implementation of $P$ which is not globally optimal as it
is not a mechanistic improvement of implementation
$Y={\Nt a; \Fj 3; X}$ of $P$.

This is worked out as follows.
Find that $\mextr{X}=Q$ defined by
\[ Q = \pcc{a}{\delay(Q)}{b\circ\st}, 
\]
and that $\mextr{Y}=Q'$, where
\[ Q' = \pcc{a}{Q}{\delay(b\circ\st)}.
\]
Notice that neither $Q\mleq Q'$ nor $Q'\mleq Q$.
\end{example}

\begin{fact}
If a thread has implementations it need 
not have a globally optimal implementation.
For example, consider again the thread
\[ P = \pcc a{b\circ\st}{c\circ\st}
\]
from the example of Fact~\ref{fact:xyz}. 
Both
\[
  X = \Pt{a};\Fj 3; c;\Tin;b;\Tin
  \quad\text{and}\quad
  Y = \Nt{a};\Fj 3; b;\Tin;c;\Tin
\]
are implementations of $P$
but they are not comparable:
\[
  \mextr X = \pcc{a}{\delay(b\circ\st)}{c\circ\st}
\]
and
\[
  \mextr Y = \pcc{a}{b\circ\st}{\delay(c\circ\st)},
\]
so that neither
$\mextr X\mleq \mextr Y$ nor 
$\mextr Y\mleq \mextr X$.
Furthermore, neither sequence can be improved, as seen
in the example of Fact~\ref{fact:xyz},
so both are optimal implementations.
\end{fact}

\begin{fact}
Every regular (that is, finite-state, see~\cite{PZ06})
thread has an implementation.
\end{fact}
Proof: For any regular thread $P$ exists
a sequence $X$ with $\extr X = P$ (see~\cite{PZ06}).
For this $X$ it holds that $P\mleq\mextr X$.

\section{Optimization of Implementations}
Optimization of an implementation concerns
the following question:
given an implementation
can we find an improved implementation
of the same behavior?
We restrict to two observations, where
the second one requires a bit more argumentation:
\begin{enumerate}
\item Implementations are improved by the unchaining of jumps.
\item Some implementable threads do not have an optimal implementation.
\end{enumerate}

We start with the first observation.
If an instruction sequence contains jumps to jump instructions,
we speak of \emph{chained\/} jumps.
For example, consider sequence
\[ X = \Fj 2; a; \Fj 1; b;\Tin
\]
where the first instruction is a jump to
the third instruction,
which is a jump to the fourth instruction.
Unchaining of jumps in this case
simply means that we jump to the target location
of the latter jump directly.
This gives
\[ X' = \Fj 3; a; \Fj 1; b;\Tin.
\]
Notice that $X'$ is a mechanistic improvement of $X$.
In~\cite{BL02}
so-called structural congruence equations are used to 
capture various cases of jump chaining.
Importantly, it is always possible to derive
a sequence without chained jumps, and
this unchaining leads to the
mechanistic improvement of sequences.
As a consequence we find this:
\begin{quote}
Any sequence $X$
can be improved to a sequence
$X'$,
i.e., with
\[ \mextr{X'}\mleq \mextr{X},
\]
such that $\mextr{X'}$ does not contain
multiple consecutive delays,
that is, $\mextr{X'}$ does not have residuals
of the form $\delay^2(Q)$.
\end{quote}
Proof idea: a multiple consecutive delay can
only result from a jump to a jump instruction,
which can always be unchained (leading to larger jumps).

\medskip

We turn to our second observation.
Consider thread $P$ defined by 
\[
  P = \pcc a{P}{Q}
  \quad\text{with}\quad
  Q = \pcc b{Q}{\st}.
\]
We demonstrate that each implementation of $P$ can 
be improved. Stated differently:

\begin{fact}
Thread $P$ has no optimal implementation.
\end{fact}
To begin with we consider $P$'s implementation
\[
  X = \Rep {\Pt a ; \Fj 4; \Pt b; \Fj 4;\Tin}.
\]
This $X$ is not an optimal implementation of $P$
because it is improved by
\[
  Y = \Rep {\Pt a ; \Fj 6; \Nt b; \Tin ; \Pt b; \Fj 4;\Tin}.
\]
To find an improvement of $Y$ one duplicates the
repeating part of $Y$: 
\[ Y = \Rep {\Pt a ; \Fj 6; \Nt b; \Tin ; \Pt b;
       \Fj 4;\Tin ; \Pt a ; \Fj 6; \Nt b; \Tin ;
       \Pt b; \Fj 4;\Tin} .
\]
Now consider
\[ Z = \Rep {\Pt a ; \Fj 8; \Nt b; \Tin ; \Nt b; \Tin ;
       \Nt b ; \Tin ; \Fj 6;\Tin ; \Pt a ; \Fj 6; \Nt b; \Tin ;
       \Pt b; \Fj 4;\Tin}
\]
and notice that $Z$ improves $Y$.

Now the proof consists of an extensive case distinction leading 
to the conclusion that a rewrite similar to the transformations from 
$X$ to $Y$ and from $Y$ to $Z$ is possible for any implementation 
$X'$ of $P$.
In particular the following facts can be obtained each 
with simple arguments most of which we leave to the reader.

\begin{enumerate}
\item An implementation $X$ can be assumed to have been 
written in such a form that,
(i) no jump leads to a termination instruction,
(ii) no jump leads to another jump (no chained jumps),
each instruction is accessible (that is there exists a
run which executes that instruction),
(iii) each occurrence of $a$ and $b$ is within either
a positive or a negative test.

\item If $X$ contains consecutive instructions $u$ and $v$
at least one of 
these is either a termination instruction or a jump.

\item Every occurrence of $b$ is either in a subsequence
$\Nt b; \Tin; u$ 
with $u$ either $\Pt b$ or $\Nt b$,
or in a subsequence
$\Pt b;\Fj k; \Tin$ for some $k > 1$.
To see this first notice 
that after a positive reply on $b$
another $b$ must be performed. Further notice that 
a subsequence of the form $\Nt b; \Tin; \Fj k$
can be rewritten as $\Pt b; \Fj{k+1}; \Tin$,
as there are no chained jumps which 
make use of the jump in $\Nt b; \Tin; \Fj k$.

\item There is at least one occurrence of $b$ in a subsequence
of the form $\Pt b;\Fj k ; \Tin$.
(Otherwise the instruction sequence cannot be written
as a finite PGA expression).
In addition it can be assumed that this occurrence
is contained in the repeating part of $X$.

\item Using the fact that $\Rep X = \Rep {X;X}$
it can be ensured that in this
subsequence $\Pt b;\Fj k; \Tin$
the jump $\Fj k$ leads to an instruction $u$
containing $b$ and moreover such that $u$ occurs within the
repeating part subsequent to the fragment
$\Pt b;\Fj k; \Tin$ that we consider.
Moreover it can be ensured that after execution of $u$
with a positive reply the next execution of $b$ is in
instruction $v$ which is also included in the repeating part
of the expression at a higher position.

\item Assuming $\Pt b;\Fj k; \Tin$, $u$ and $v$ as
in the previous item, two 
cases are now distinguished: $u = \Pt b$ and $u = \Nt b$.
In the first case $u$ is followed by $\Fj l; \Tin$ for some
positive $l$ with $\Fj l$ leading to $v$,
while in the second case $u$ is followed by $\Tin; v$
or by $\Tin; \Fj m$ with $\Fj m$ leading to $v$.

In the first case an improvement of $X$ is found by replacing the
subsequence
$\Pt b; \Fj k; \Tin$ by
$\Nt b; \Tin ; \Pt b;\Fj {k'}; \Tin $ with
jump $\Fj {k'}$ leading to $v$.
In this case all jumps that `fly over' the
modified part need to be increased by 2.
The second case has two subcases: if $u$ is followed by
$\Tin; \Fj m$ an
improvement is found by replacing $\Pt b;\Fj k; \Tin$
by $\Nt b; \Tin; \Pt b;\Fj {k'}; \Tin$ again
with jump $\Fj {k'}$ 
leading to $v$, and while appropriately increasing other jumps
in the instruction sequence. 

\item We are left with the remaining case that $u$ is followed
by $\Tin; v$.
Now observe that if we consider subsequent instructions it cannot
be an indefinite repetition of $\Nt b; \Tin$ and at some stage
a either a positive test followed by a jump ($\Pt b$) or
a jump following termination must occur.
We consider one such case only, the 
other variations being dealt with similarly.
Let $u$ be the start of a subsequence
$\Nt b; \Tin; \Nt b; \Tin; \Nt b; \Tin; \Pt b; \Fj n; \Tin$.
Then an improvement is 
found by expanding $\Pt b;\Fj k; \Tin$ to
$\Nt b; \Tin;\Nt b; \Tin; \Nt b; \Tin; \Nt b; \Tin; \Pt b; \Fj {n'}; \Tin$,
with $n'$ chosen in such a way that it leads to $v$, and
increasing all jumps that `jump over' the expanded part of the
instruction sequence by $6$.
\end{enumerate}

\section{Concluding Remarks}
Mechanistic thread extraction preserves some 
information concerning the 
computational mechanisms invoked 
by an instruction sequence. 
Our result that the thread
$\extr{\Rep {\Pt a; \Fj 4; \Pt b; \Fj 4; \Tin}}$ 
has no optimal implementation 
suggests that improvements are 
possible for each implementation. 
Such improvements give rise to instruction sequences
with increasingly longer repeating parts.
This implies a decrease in code compactness which,
somehow, will eventually lead to slower computations.
Balancing code compactness 
versus improved implementation cannot 
be done in the absence of numerical data on 
implementation technologies and for that
reason no attempt is made to do so here.

We have defined and studied mechanistic thread extraction 
for the simplest of program notations in the program algebra family
as presented in~\cite{BL02}. 
For each new instruction
one may provide a mechanistic thread extraction policy.
In defining mechanistic behavior of
instruction sequences there are 
several degrees of freedom.
For instance one might 
insist that an absolute jump requires
a single unit of time,
whereas a relative jump takes two, 
in view of the fact that performing
a relative jump requires some 
arithmetic involving the program counter. 

Such decisions are to some extent arbitrary and it 
should be expected that in specific cases
the most useful definition of mechanistic behavior 
of a PGA instruction sequence may differ
from what we have defined above.
For instance, jumps with a counter 
exceeding some large value, e.g., 100000,
may be assigned a larger cost than small jumps. 
This modification already may 
invalidate the result that
some finite state threads have no optimal 
implementation (for the particular
definition of mechanistic thread extraction 
as given above). 

Many instructions can be analyzed in mechanistic 
terms, we mention for instance:
backward jumps, absolute jumps, goto's,
indirect jumps (of various kind), 
returning jumps, calls to a service, 
calls to a blocking service,
instructions that cause thread creation or thread migration, 
calls to another instruction sequence and unit instructions.

Mechanistic behavior has a focus on the numbers of steps 
needed for an immediate interpretation
of an instruction sequence.
This is by no means the only conceivable cost factor.
Modeling energy consumption may be just as important and if 
basic actions are measured concerning their cost of 
execution the avoidance of expensive (with respect to time or 
energy or risk of failure) actions in favor of cheap 
ones may be more important than the 
minimization of the number of jumps.

\end{document}